\documentclass[journal,twoside,web]{ieeecolor}

\usepackage{generic}
\usepackage{lcsys}
\usepackage{amsmath,amssymb,amsfonts}
\usepackage{booktabs}
\usepackage{adjustbox}
\usepackage{wrapfig}
\usepackage{lipsum}
\usepackage{amssymb}
\usepackage{amsmath}
\usepackage{siunitx}
\usepackage{mathtools}
\usepackage{hyperref}
\let\labelindent\relax
\usepackage[inline]{enumitem}
\usepackage[ruled,vlined,linesnumbered]{algorithm2e}
\usepackage[noend]{algpseudocode}
\usepackage{ifthen}
\usepackage{dsfont}
\usepackage[backend=biber,style=numeric-comp,sorting=none,maxbibnames=99]{biblatex}
\usepackage{balance}
\usepackage{dirtytalk}
\usepackage{afterpage}
\usepackage{bm}

\usepackage{times}

\allowdisplaybreaks

\newcommand{\G}{\mathcal{G}}

\newcommand{\ra}{\rightarrow}
\newcommand{\R}{\mathbb{R}}

\newtheorem{remark}{Remark}
\newtheorem{assumption}{Assumption}
\newtheorem{proposition}{Proposition}
\newtheorem{definition}{Definition}
\newtheorem{lemma}{Lemma}
\newtheorem{theorem}{Theorem}

\newtheorem{example}{Example}

\newcommand{\height}{m}

\newcommand{\revision}[1]{\textcolor[rgb]{0,0,0}{#1}}

\usepackage{todonotes}
\usepackage{bbm}

\bibliography{references}

\def\BibTeX{{\rm B\kern-.05em{\sc i\kern-.025em b}\kern-.08em
T\kern-.1667em\lower.7ex\hbox{E}\kern-.125emX}}

\begin{document}
\title{To What Extent do Open-loop and Feedback Nash Equilibria Diverge in General-Sum Linear Quadratic Dynamic Games?}

\author{Chih-Yuan Chiu$^{\star}$, \IEEEmembership{Member, IEEE}, Jingqi Li$^{\star}$, \IEEEmembership{Student Member, IEEE}, Maulik Bhatt and Negar Mehr, \IEEEmembership{Member, IEEE}
\thanks{$^\star$Equal contribution}
\thanks{$^{1}$Chih-Yuan Chiu is with the School of Electrical and Computer Engineering, Georgia Institute of Technology, Atlanta, GA, USA (\texttt{cyc at gatech dot edu}).}
\thanks{$^{2}$Jingqi Li is with the Department of Electrical Engineering and Computer Sciences, University of California, Berkeley, CA, USA (\texttt{jingqili at berkeley dot edu}).}
\thanks{$^{3}$Maulik Bhatt and Negar Mehr are with the Department of Mechanical Engineering, University of California, Berkeley, CA, USA (\texttt{\{maulikbhatt, negar\} at berkeley dot edu}) }
\thanks{This work is supported by the National Science Foundation grants ECCS-2145134 CAREER Award, CNS-2423130, and CCF-2423131.}
}

\maketitle
\thispagestyle{empty}

\begin{abstract}
Dynamic games offer a versatile framework for modeling the evolving interactions of strategic agents, whose steady-state behavior can be captured by the \emph{Nash equilibria} of the games. Nash equilibria are often computed in \textit{feedback}, with policies depending on the state at each time, or in \textit{open-loop}, with policies depending only on the initial state. Empirically, open-loop Nash equilibria (OLNE) could be more efficient to compute, while feedback Nash equilibria (FBNE) often encode more complex interactions. However, it remains unclear exactly which dynamic games yield FBNE and OLNE that differ significantly and which do not. To address this problem, we present a principled comparison study of OLNE and FBNE in linear quadratic (LQ) dynamic games. Specifically, we prove that the OLNE strategies of an LQ dynamic game can be synthesized by solving the coupled Riccati equations of an auxiliary LQ game with perturbed costs. The construction of the auxiliary game allows us to establish conditions under which OLNE and FBNE coincide and derive an upper bound on the deviation between FBNE and OLNE of an LQ game.

\end{abstract}


\section{Introduction}
\label{sec: Introduction}


Dynamic game theory \cite{basar1998DynamicNoncooperativeGameTheory} provides a powerful mathematical framework for modeling strategic decision-making over time in multi-agent interactions, such as autonomous path planning \cite{vinitsky2023optimizing, bhatt2023efficient, cleac2019algames, fridovich2020efficient}.
In dynamic games, steady-state modes of interaction between strategic agents are described by \textit{Nash equilibria} \cite{nash1950equilibrium,basar1998DynamicNoncooperativeGameTheory}, strategy profiles at which each agent's behavior is unilaterally optimal with respect to their objective. Different Nash equilibria concepts can be defined with respect to different \textit{information structures}, which prescribe the system state information accessible to agents for control design at each time, and thus influence their strategic interactive behavior. 

In \cite{starr1969nonzero,lowe2017multi,fridovich2020efficient, laine2021computation, hambly2023policy, Monti2024FeedbackOpenLoopNashEquilibriaInfiniteHorizon, Benenati2024LinearQuadraticDynamicGamesRecedingHorizon}, iterative algorithms were developed for computing Nash equilibria under the \textit{feedback} information structure, wherein each agent observes the current system state at each time and adapts their actions accordingly. Other works solve for Nash equilibria 
in \textit{open-loop}
, in which each agent selects control actions using only the state information available at the start of the planning horizon 
\cite{starr1969nonzero,reddy2015open,cleac2019algames, bhatt2023efficient, Monti2024FeedbackOpenLoopNashEquilibriaInfiniteHorizon, Benenati2024LinearQuadraticDynamicGamesRecedingHorizon}.

Prior works \cite{basar1998DynamicNoncooperativeGameTheory,laine2021computation,Li2023CostInferenceforFeedbackDynamicGames} suggest that the feedback Nash equilibrium (FBNE) and open-loop Nash equilibrium (OLNE) trajectories of some dynamic games can diverge wildly, even when the OLNE are generated in a receding horizon fashion  \cite{fridovich2020efficient, Li2023CostInferenceforFeedbackDynamicGames}. Moreover, in dynamic games whose FBNE and OLNE diverge significantly, FBNE often describes nuanced multi-agent interactions more accurately than OLNE \cite{fridovich2020efficient, laine2021computation}. However, some studies indicate that OLNE is more computationally tractable to solve than FBNE, and that FBNE and OLNE are closely aligned for many dynamic games \cite{bhatt2023efficient}. These empirical phenomena lead naturally to the following question: \textit{For which dynamic games do FBNE diverge sufficiently from OLNE to warrant the computational burden of computing FBNE?}

To answer the above question, we characterize the difference between the FBNE and OLNE solutions, \revision{henceforth referred to as the FBNE-to-OLNE gap}, for a class of linear quadratic (LQ) dynamic games. First, we identify sufficient conditions for LQ dynamic games under which their FBNE and OLNE coincide. Then, for LQ games whose FBNE and OLNE solutions diverge, we establish an upper bound for their FBNE-to-OLNE gap in terms of the degree to which each agent's cost depends on other agents' state values. Our core method involves proving that the OLNE of any LQ game $\G$ satisfying certain structural properties can be synthesized by constructing an \emph{auxiliary} game $\tilde{\G}$ with modified cost matrices and solving the Riccati equations of $\tilde{\G}$. 
The construction of the auxiliary game allows us to recast the comparison between the OLNE and FBNE of $\G$ as a comparison between the 
Riccati equation solutions for
$\G$ and $\tilde \G$, which we then characterize by contrasting the state costs of $\G$ and~$\tilde \G$. 

Below, Section \ref{sec: Preliminaries} presents the class of 
LQ games considered in this work, and describes the FBNE and OLNE solutions of LQ Games. Section \ref{sec: Sufficient Conditions for Alignment of FBNE and OLNE Trajectories} then presents sufficient conditions under which the FBNE and OLNE of a given LQ game coincide. Section \ref{sec: Bounding the Gap Between the FBNE and OLNE of LQ Games} 
upper bounds
the FBNE-to-OLNE gap of LQ games in terms of their state cost matrices. Section \ref{sec: Simulation Studies} presents numerical simulations supporting our theoretical~results.

\textit{Notation}:
For each positive integer $n$, set $[n]:=\{\\\allowdisplaybreaks 1, \cdots, n\}$. Given matrices $(A^1, \cdots, A^n)$, let $\text{diag}\{A^1, \cdots, \\\allowdisplaybreaks A^n\}$ be the block diagonal matrix whose $i$-th block is $A^i$, $\forall i \in [n]$. 
Given a matrix $M$ with a block structure, let $[M]_{ij}$, $[M]_{i, :}$, and $[M]_{:,j}$  respectively denote the $i$-th row $j$-th column block, the $i$-th row block (with all columns), and the $j$-th column block (with all rows) of $M$. Let $I$ and $O$ denote the identity and zero matrix, respectively.

\section{Preliminaries}
\label{sec: Preliminaries}

\subsection{Linear Quadratic Games}
\label{subsec: Setup}

We denote by $\G := \big( A^i, B^i, Q^i, R^i: i \in [N] \big)$ an $N$-agent, $T$-stage discrete-time dynamic LQ game with structure described as follows. Each agent $i \in [N]$ is associated with the states $x_t^i \in \R^{n_i}$ and controls $u_t^i \in \R^{m_i}$ $\forall \ t \in [T]$, which evolve according to the dynamics:\vspace{-0.5em}
\begin{equation} \label{Eqn: System Dynamics, Each agent}
    \Sigma_i: \ x_{t+1}^i = A^i x_t^i + B^i u_t^i, \ \forall \ t \in [T],\vspace{-0.3em}
\end{equation}
where $A^i \in \R^{n_i \times n_i}$ and $B^i \in \R^{n_i \times m_i}$. We define $x_t := (x_t^1, \cdots, x_t^N) \in \R^n$ to be the \textit{system state} at each time $t \in [T]$, where $n := \sum_{i \in [N]} n_i$, and $\textbf{x} := (x_1, \cdots, x_T) \in \R^{nT}$ to be the \textit{system state trajectory}. The initial state $x_1$ is assumed to be fixed. Similarly, we define $u_t := (u_t^1, \cdots, u_t^N) \in \R^m$ to be the \textit{system control input} at time $t \in [T]$, where $m := \sum_{i \in [N]} m_i$.  We denote each agent $i$'s control inputs over time by $u^i := (u_1^i, \cdots, u_T^i) \in \R^{m_i T}$, and the tuple of all agents' control inputs over time by $u := (u^1, \cdots, u^N) \in \R^{mT}$. Each agent $i \in [N]$ is associated with the cost function $C^i: \R^{nT} \times \R^{m_iT} \ra \R$:
\vspace{-0.5em}
\begin{equation} \label{Eqn: Cost, Each agent}
    C^i(\textbf{x}, u^i) := \sum_{t\in[T]} \big[ u_t^{i\top} R^i u_t^i + x_{t+1}^\top Q^i x_{t+1} \big].\vspace{-0.5em}
\end{equation}
where $R^i \in \R^{m_i \times m_i}$ and $Q^i \in \R^{n \times n}$ denote the control cost matrices and state cost matrices of agent $i$, respectively.

\begin{remark}
In this work, we focus our analysis on LQ games with \textit{decoupled} dynamics \eqref{Eqn: System Dynamics, Each agent} and \textit{coupled} state costs \eqref{Eqn: Cost, Each agent}. Thus, the state cost of each agent $i$ may depend on the states of other agents, but their control cost and the dynamics depend only on their own control $u_t^i$. This assumption reasonably describes many multi-agent interaction applications, such as autonomous driving and multi-robot coordination.
\end{remark} 


\begin{remark}
The terms $x_{t+1}^\top Q^i x_{t+1}$ in \eqref{Eqn: Cost, Each agent} encode the dependence of each agent's cost on the entire state trajectory, thus enforcing that the agents' costs are coupled.
\end{remark}

System-level dynamics for the game $\G$ can be constructed by concatenating the system dynamics $\Sigma_i$ in \eqref{Eqn: System Dynamics, Each agent}, as follows:\vspace{-0.3em}
\begin{equation} 
    \Sigma: \ x_{t+1} = A x_t + B u_t  
    = A x_t + \sum_{i \in [N]} \hat B^i u_t^i, \label{Eqn: System Dynamics, N agents, with B matrices per agent} \vspace{-0.5em}
\end{equation}
where 
we have $A := \text{diag}\{A^1, \cdots, A^N \} \in \R^{n \times n}$, $B := \text{diag}\{B^1, \cdots, B^N \} \in \R^{n \times m}$. 
We define $\hat B^i$ to be the $i$-th block columns of $B$, i.e., $\hat B^i := [B]_{:, i}$ $\forall \ i \in [N]$.


\subsection{Feedback Nash Equilibria (FBNE)}
\label{subsec: Feedback Nash Equilibria (FBNE)}

The state feedback information structure prescribes that, at each time $t$, each agent is allowed to access the value of the system state $x_t \in \R^n$ to guide the selection of their current control input $u_t^i \in \R^{m_i}$. Thus, each agent $i$ designs \textit{feedback strategies} $\gamma_t^i: \R^n \ra \R^{m_i}$ over the time horizon $[T]$ a priori, and subsequently deploys the control input $u_t^i := \gamma_t^i(x_t)$ at each time $t$ after obtaining the value of $x_t$. We define $\Gamma_t^i$ to be the strategy space of each agent $i$ at each time $t$, and $\Gamma$ to be the joint strategy space of all agents over the time horizon. Given a tuple of state feedback strategies $\gamma \in \Gamma$, for each agent $i \in [N]$, for any $\tau, \bar \tau \in [T]$, define $\gamma_{\tau:\bar \tau}^i$ to be $(\gamma_\tau^i, \cdots, \gamma_{\bar \tau}^i)$ if $\tau < \bar \tau$, and empty otherwise. Finally, for each agent $i \in [N]$, let the notation \say{$-i$} denote \say{all agents except agent $i$}, e.g., for any $\tau, \bar \tau \in [T]$ such that $\tau < \bar \tau$, we have $\gamma_{\tau, \bar \tau}^{-i} := (\gamma_t^j \in \Gamma_t^j: \tau \leq t \leq \bar\tau, j \in [N] \backslash \{i\})$.

Below, we define the Nash equilibrium of a dynamic game under the state feedback information structure using notation introduced in~\cite{basar1998DynamicNoncooperativeGameTheory}. 
For each $i \in [N]$, let $J^{i, FB}: \Gamma \ra \R$ denote the cost obtained by unrolling the system dynamics \eqref{Eqn: System Dynamics, Each agent} from the fixed initial state $x_1$ with control inputs prescribed by a given state feedback strategy, i.e., for each feedback strategy $\gamma \in \Gamma$, we have:\vspace{-0.2em}
{ 
\begin{align*}
    J^{i, FB}(\gamma) := C^i(\textbf{x}, \gamma_1^i(x_1), \cdots, \gamma_T^i(x_T) )
\end{align*}
}where $\textbf{x}$ satisfies \eqref{Eqn: System Dynamics, Each agent} with $u_t^i = \gamma_t^i(x_t)$ $\forall \ i \in [N], t \in [T]$.


\begin{definition}[\textbf{Feedback Nash Equilibrium (FBNE)}] (\cite{basar1998DynamicNoncooperativeGameTheory}, Chapter 6) \label{Def: FBNE}
The strategy tuple $\gamma^\star \in \Gamma$ is called a \textit{feedback Nash equilibrium} (FBNE) if it is unilaterally optimal, i.e., if for all $\gamma \in \Gamma$, at each $i \in [N]$ and $t \in [T]$:\vspace{-0.2em}
{
\begin{align*}
    &J^{i, FB}(\gamma_{1:t-1}^i, \textcolor{red}{\bm \gamma_t^{i\star}}, \gamma_{t+1:T}^{i\star}; \gamma_{1:t-1}^{-i}, \gamma_t^{-i\star}, \gamma_{t+1:T}^{-i\star}) \\
    \leq \ &J^{i, FB}(\gamma_{1:t-1}^i, \textcolor{red}{\bm \gamma_t^i}, \gamma_{t+1:T}^{i\star}; \gamma_{1:t-1}^{-i}, \gamma_t^{-i\star}, \gamma_{t+1:T}^{-i\star}).
\end{align*}
}
\end{definition}

Def. \ref{Def: FBNE} imposes \textit{strong time consistency} (\cite{basar1998DynamicNoncooperativeGameTheory}, Def. 5.14), i.e., a FBNE is unilaterally optimal for any subgame starting at any intermediate stage $t \in [T]$ from any state.

It is well known that the FBNE of LQ games with structure described in Section \ref{subsec: Setup} can be characterized in closed-form \cite{basar1998DynamicNoncooperativeGameTheory}. Let $\G = \big(A^i, B^i, Q^i, R^i: {i \in [N]} \big)$ denote an LQ game of the form described in Section \ref{subsec: Setup}. Then $Z_t^i \in \R^{n \times n}$, $K_t^i \in \R^{m_i \times n}$, and $F_t \in \R^{n \times n}$ $\forall \ i \in [N]$, $t \in [T]$ can be \textit{recursively computed backwards in time from $t = T$} using the following coupled Riccati equations:\vspace{-0.4em}
\begin{align} \label{Eqn: FBNE, Riccati, 1, initialization}
    &Z_{T+1}^i = Q^i, \\ \label{Eqn: FBNE, Riccati, 2, computing Kt}
    &R^i K_t^i + \hat B^{i\top} Z_{t+1}^i \sum_{\revision{j} \in [N]} \hat B^j K_t^j = \hat B^{i\top} Z_{t+1}^i \revision{A}, \\ \label{Eqn: FBNE, Riccati, 3, computing Ft}
    &F_t = A - \sum_{\revision{j} \in [N]} \hat B^{\revision{j}} K_t^{\revision{j}}, \\ \label{Eqn: FBNE, Riccati, 4, computing Zt}
    &Z_t^i = Q^i + F_t^\top Z_{t+1}^i F_t + {K_t^i}^\top R^i K_t^i.
\end{align}
\revision{In particular, \eqref{Eqn: FBNE, Riccati, 1, initialization} provides boundary conditions for $(Z_t^i: i \in [N])$ at $t = T+1$. Then, given $(Z_{t+1}^i: i \in [N])$ at some time $t$, \eqref{Eqn: FBNE, Riccati, 2, computing Kt} provides a system of linear equations for computing $K_t^i$. Subsequently, \eqref{Eqn: FBNE, Riccati, 3, computing Ft} computes $F_t$ using $K_t^i$, and \eqref{Eqn: FBNE, Riccati, 4, computing Zt} computes $Z_t^i$ using $F_t$ and $K_t^i$. We then decrement $t$ by 1 and return to \eqref{Eqn: FBNE, Riccati, 2, computing Kt} to solve for $(K_{t-1}^i: i \in [N])$.}

For convenience, let $P_t \in \R^{m \times m}$ and $K_t \in \R^{m \times n}$, $S_t \in \R^{m \times n}$ be block-wise defined such that $[P_t]_{ii} = R^i + B^{i\top} Z_{t+1}^i B^i$ and $[P_t]_{ij} = B^{i\top} Z_{t+1}^i B^j$, $[K_t]_{i, :} := K_t^i$, and $[S_t]_{i, :} := B^{i\top} Z_{t+1}^i A$, $\forall \ i, j \in [N]$, $i \ne j$. Then \eqref{Eqn: FBNE, Riccati, 2, computing Kt} can be rewritten compactly as:\vspace{-0.4em}
\begin{equation} \label{Eqn: Compact version of FBNE, Riccati, 2, computing Kt}
    P_t K_t = S_t.\vspace{-0.4em}
\end{equation}

Below, we present Assumption \ref{Assumption: LQ Games}, which is maintained throughout the paper to ensure that the dynamic LQ games defined above yield a unique FBNE~(\cite{basar1998DynamicNoncooperativeGameTheory},~Remark 6.4). 

\begin{assumption} \label{Assumption: LQ Games}
$Q^i$ is positive semi-definite and $R^i$ is positive definite $\forall \ i \in [N]$. Moreover, $P_t^{-1}$ exists $\forall \ t \in [T]$.
\end{assumption}

Assumption \ref{Assumption: LQ Games} ensures that the coupled Riccati equations have a unique solution,~as~stated~in~the~following~proposition. 
\begin{proposition}[\cite{basar1998DynamicNoncooperativeGameTheory}, Corollary 6.1] \label{Prop: FBNE, Riccati}
If a dynamic LQ game $\G = \big(A^i, B^i, Q^i, R^i:i \in [N]\big)$ satisfies Assumption \ref{Assumption: LQ Games}, it admits unique FBNE strategies and trajectory  
given as follows, $\forall \ i \in [N]$, $t \in [T]$:\vspace{-0.6em}
\begin{align} \label{Eqn: FBNE Strategy, from Basar}
    \gamma_t^{i^\star}(x_t) &= - K_t^i x_t, \\ \label{Eqn: FBNE Trajectory, from Basar}
    x_{t+1} &= F_t x_t.
\end{align}

\end{proposition}\vspace{-0.3em}

\subsection{Open-Loop Nash Equilibria (OLNE)}
\label{subsec]: Open-Loop Nash Equilibria (OLNE)}

In contrast to the state feedback information structure, the \textit{open-loop information structure} requires each agent $i$ to \textit{only} use the initial state $x_1 \in \R^n$ to design their control $u_t^i$ at each time $t$.  In this case, each agent $i$ devises \textit{open-loop} strategies $\phi_t^i: \R^n \ra \R^{m_i}$ over the time horizon $[T]$ a priori, and then deploys the control input $u_t^i = \phi_t^i(x_1)$ at each time $t$. We define $\Phi$ to be the open-loop joint strategy space of all agents. 


Below, we define the Nash equilibrium corresponding to the open-loop information structure. For each agent $i \in [N]$, let $J^{i, OL}: \Phi \ra \R$ denote the cost obtained by unrolling the system dynamics \eqref{Eqn: System Dynamics, Each agent} with control inputs prescribed by a given open-loop strategy, i.e., 
{
\begin{align*}
    J^{i, OL}(\phi) := C^i(\textbf{x}, \phi_1^i(x_1), \cdots, \phi_T^i(x_1))
\end{align*}
}where $\textbf{x}$ satisfies \eqref{Eqn: System Dynamics, Each agent} with $u_t^i = \phi_t^i(x_1)$ $\forall \ i \in [N]$,~$ t \in [T]$. 

\begin{definition}[\textbf{Open-Loop Nash Equilibrium (OLNE)}] \label{Def: OLNE}(\cite{basar1998DynamicNoncooperativeGameTheory}, Chapter 6) 
We call $(u^{1\star}, \cdots, u^{N\star}) \in \R^{mT}$ an \textit{open-loop Nash Equilibrium} (OLNE) if for any open-loop strategy $\phi \in \Phi$, and agent $\forall \ i \in [N]$, we have:
{
\begin{align*}
    J^{i, OL}(\textcolor{red}{\bm \phi^{i, \star}}; \phi^{-i,\star}) \leq J^{i, OL}(\textcolor{red}{\bm \phi^i}; \phi^{-i,\star}).
\end{align*}
}
\end{definition}

The OLNE of LQ games with the structure described in Section \ref{subsec: Setup} can likewise be characterized in closed-form \cite{basar1998DynamicNoncooperativeGameTheory}. Let $\Lambda_t^i \in \R^{n \times n}$, $M_t^i \in \R^{m_i \times n}$, and $L_t^i \in \R^{n \times n}$ $\forall \ i \in [N]$, $t \in [T]$ be recursively computed backwards in time from $t = T$ using the following Riccati equations:\vspace{-0.3em}
\begin{align} \label{Eqn: OLNE, Riccati, 1, initialization}
    &M_{T+1}^i = Q^i, \\ \label{Eqn: OLNE, Riccati, 2, computing Lambda t}
    &\Lambda_t = I + \sum_{i\in[N]} \hat B^i (R^i)^{-1} \hat B^{i\top} M_{t+1}^i, \\ \label{Eqn: OLNE, Riccati, 3, computing Lt}
    &L_t^i = (R^i)^{-1} \hat B^{i\top} M_{t+1}^i \Lambda_t^{-1} A, \\ \label{Eqn: OLNE, Riccati, 4, computing Mt}
    &M_t^i = Q^i + A^\top M_{t+1}^{i} \Lambda_t^{-1} A.
\end{align}

\begin{proposition}[\cite{basar1998DynamicNoncooperativeGameTheory}, Thm. 6.2] \label{Prop: OLNE, Riccati}
Under Assumption~\ref{Assumption: LQ Games}, an LQ game $\G := \big(A^i, B^i, Q^i, R^i:i \in [N] \big)$ admits a unique OLNE if $\Lambda_t^{-1}$ exists $\forall \ t \in [T]$ and \eqref{Eqn: OLNE, Riccati, 1, initialization}-\eqref{Eqn: OLNE, Riccati, 4, computing Mt} admit a unique solution, in which case the OLNE strategies and trajectory are, $\forall \ i \in [N]$,~$t \in [T]$:\vspace{-0.3em}
\begin{align} \label{Eqn: OLNE Strategy, from Basar}
    \phi_t^{i^\star}(x_1) &= - L_t^i x_t = -L_t^i \cdot \Pi_{\tau=1}^{t-1} (\Lambda_\tau^{-1}A)x_1, \\ \label{Eqn: OLNE Trajectory, from Basar}
    x_{t+1} &= \Lambda_t^{-1} A x_t.
\end{align}
\end{proposition}

\revision{Throughout the rest of this paper, we assume that Assumption 1 holds, and that $\Lambda_t^{-1}$ exists for all $t \in [T]$.}

\section{Sufficient Conditions for Alignment of FBNE and OLNE Trajectories}
\label{sec: Sufficient Conditions for Alignment of FBNE and OLNE Trajectories}

Empirical studies suggest that FBNE and OLNE strategies are more closely aligned when the state cost matrix of each agent has minimal dependence on the states and inputs of other agents. In Section \ref{subsec: Auxiliary Game}, we explicitly quantify this observation by associating a given LQ game $\G := (A^i, B^i, Q^i, R^i: i \in [N])$ with an \textit{auxiliary} game $\tilde \G := (A^i, B^i, \tilde Q^i, R^i: i \in [N])$. The auxiliary game $\tilde \G$ is a copy of the original LQ game $\G$ whose state cost matrices $\tilde Q^i$ have been modified, to reduce the dependency of each agent's cost on other agents' states. We then 
use auxiliary games to contrast the FBNE and OLNE solutions for general-sum LQ dynamic games in Section \ref{subsec: FBNE and OLNE of LQ Games and their Auxiliary Games}.\vspace{-0.4em}


\subsection{Auxiliary Game}
\label{subsec: Auxiliary Game}

Formally, for a given LQ dynamic game, we define our auxiliary game as follows.

\begin{definition}[\textbf{Auxiliary LQ Game}]
Given an LQ Game $\G := \big( A^i, B^i, Q^i, R^i: i \in [N] \big)$, we define the \textbf{auxiliary game} associated with $\G$ to be $\tilde \G := \big( A^i, B^i, \tilde Q^i, R^i: i \in [N] \big)$, where $\tilde Q^i$ is defined by retaining the elements of $Q^i$ in the $i$-th row block while setting all other elements to 0, i.e., $[\tilde Q^i]_{i, :} = [Q^i]_{i, :}$ and $[\tilde Q^i]_{j, :} = O$ $\forall \ j \ne i$.
\end{definition}

\begin{remark}
Auxiliary LQ games in general do not have symmetric $Q^i$, and thus are not LQ games and do not have equilibria in the conventional sense. Rather, auxiliary games are artificial constructs defined to assist us in establishing algebraic conditions under which the FBNE and OLNE trajectories of an LQ game are aligned. Nonetheless, for ease of exposition, we will refer to matrices satisfying \eqref{Eqn: FBNE, Riccati, 1, initialization}-\eqref{Eqn: Compact version of FBNE, Riccati, 2, computing Kt} for an auxiliary game $\tilde \G := \big( A^i, B^i, \tilde Q^i, R^i: i \in [N] \big)$, if they exist, as characterizing the \say{FBNE} of $\tilde \G$, and denote them by $(\tilde Z_t^i, \tilde K_t^i, \tilde F_t, \tilde P_t, \tilde S_t: i \in [N], t \in [T])$. Similarly, we will refer to matrices satisfying \eqref{Eqn: OLNE, Riccati, 1, initialization}-\eqref{Eqn: OLNE, Riccati, 4, computing Mt} for $\tilde \G$, if they exist, as characterizing the \say{OLNE} of $\tilde \G$, and denote them by $(\tilde M_t^i, \tilde \Lambda_t, \tilde L_t^i: i \in [N], t \in [T])$.
\end{remark}


\subsection{FBNE/OLNE of LQ Games vs. Auxiliary LQ Games}
\label{subsec: FBNE and OLNE of LQ Games and their Auxiliary Games}


First, we show that the OLNE of an LQ game $\G$ satisfies \eqref{Eqn: OLNE, Riccati, 1, initialization}-\eqref{Eqn: OLNE, Riccati, 4, computing Mt} for $\tilde \G$, i.e., loosely speaking, the OLNE of $\G$ and of $\tilde \G$ coincide.

\begin{lemma} \label{Lemma: 4th lemma in Jingqi notes}
Under Assumption~\ref{Assumption: LQ Games}, let $\tilde \G$ be the auxiliary game associated with $\G$. Suppose that $\Lambda_t^{-1}$ exists for all $t$, then the OLNE of $\G$ satisfies \eqref{Eqn: OLNE, Riccati, 1, initialization}-\eqref{Eqn: OLNE, Riccati, 4, computing Mt} for $\tilde \G$. \revision{In other words, $\tilde{L}_t^i = L_t^i$, $\forall i\in[N]$, $t\in[T]$.}
\end{lemma}

\begin{proof}
From \eqref{Eqn: OLNE, Riccati, 2, computing Lambda t}-\eqref{Eqn: OLNE Trajectory, from Basar}, we find it suffices to prove that 
$\hat B^{i\top} \tilde M_{t+1}^i = \hat B^{i\top} M_{t+1}^i$ $\forall \ t \in [T]$, $i \in [N]$. Due to the block structure of $\hat B^{i\top}$, it suffices to prove that $[\tilde M_{t+1}^i]_{i, :} = [M_{t+1}^i]_{i, :}$. We proceed by backward induction from time $T$. When $t = T$, we have $[\tilde M_{T+1}^i]_{i, :} = [\tilde Q^i]_{i, :} = [Q^i]_{i, :}  = [M_{t+1}^i]_{i, :}$. Suppose $[\tilde M_{t+1}^i]_{i, :} = [M_{t+1}^i]_{i, :}$ for some $t \in [T]$. From the block structures of each $\hat B^i$ and of $A$, we have:\vspace{-0.4em}
\begin{equation*}
\hat B^{i\top} \tilde M_{t+1}^i =  B^{i\top} [\tilde M_{t+1}^i]_{i,:} =  B^{i\top} [M_{t+1}^i]_{i,:} = \hat B^{i\top} M_{t+1}^i.\vspace{-0.5em}
\end{equation*}
From \eqref{Eqn: OLNE, Riccati, 2, computing Lambda t}, we obtain $\tilde \Lambda_t = \Lambda_t$, and $\tilde \Lambda_t^{-1}$ exists. Then:\vspace{-0.4em}
\begin{align} \label{Eqn: Proof of Lemma on OLNE, 1}
    [\tilde M_t^i]_{i,:} &= [\tilde Q^i]_{i, :} + [A^\top \tilde M_{t+1}^i]_{i, :} \Lambda_t^{-1} A \\ \label{Eqn: Proof of Lemma on OLNE, 2}
    &= [\tilde Q^i]_{i, :} + A^{i\top} [\tilde M_{t+1}^i]_{i, :} \Lambda_t^{-1} A \\ \label{Eqn: Proof of Lemma on OLNE, 3}
    &= [Q^i]_{i, :} + A^{i\top} [M_{t+1}^i]_{i, :} \Lambda_t^{-1} A \\ \nonumber
    &= [Q^i]_{i, :} + [A^\top M_{t+1}^i]_{i, :} \Lambda_t^{-1} A \\ \label{Eqn: Proof of Lemma on OLNE, 4}
    &= [M_t^i]_{i,:}
\end{align}
where \eqref{Eqn: Proof of Lemma on OLNE, 2} follows from the block structure of $A$, and \eqref{Eqn: Proof of Lemma on OLNE, 3} follows from the fact that  $[\tilde Q^i]_{i, :} = [Q^i]_{i, :}$, and the induction hypothesis that $[\tilde M_{t+1}^i]_{i, :} = [M_{t+1}^i]_{i, :}$. \revision{This completes the induction step and thus proves that $\tilde{L}_t^i = L_t^i$, $\forall i\in[N], t\in[T]$, and $\Lambda_t^{-1} A = \tilde \Lambda_t^{-1}A \ \forall \ t \in [T]$.}
\end{proof}

\begin{remark} \label{Remark: Mti only has nonzero ith row blocks}
From \eqref{Eqn: Proof of Lemma on OLNE, 1}-\eqref{Eqn: Proof of Lemma on OLNE, 4}, we also find that $\tilde M_{t+1}^i$ has nonzero entries only in the $i$-th row block, $\forall \ i \in [N], t \in [T]$. Indeed, $[\tilde M_{T+1}^i]_{j,:} = [\tilde Q^i]_{j,:} = O$. Thus, if $[\tilde M_{t+1}^i]_{j,:} = O, \ \forall \ j \ne i$, for some $t \in [T]$, then, we have:\vspace{-0.4em}
\begin{equation} \nonumber
\begin{aligned}
    [\tilde M_t^i]_{j,:} &= [\tilde Q^i]_{j, :} + [A^\top \tilde M_{t+1}^i]_{j, :} \Lambda_t^{-1} A \\ \nonumber
    &= O + A^{j\top} [\tilde M_{t+1}^i]_{j, :} \Lambda_t^{-1} A \\ \nonumber
    &= O, 
\end{aligned}\vspace{-0.5em}
\end{equation}
which implies that $[\tilde M_t^i]_{j,:} = O \ \forall \ j \ne i$, $t \in [T]$.
\end{remark}

Next, we prove that the OLNE solution of any auxiliary LQ Game $\tilde \G$ always satisfies the Riccati equations \eqref{Eqn: FBNE, Riccati, 1, initialization}-\eqref{Eqn: FBNE, Riccati, 4, computing Zt} for $\tilde \G$. Loosely speaking, the OLNE and FBNE of $\tilde \G$ coincide.

\begin{lemma} \label{Lemma: 3rd lemma in Jingqi notes}
Under Assumption~\ref{Assumption: LQ Games}, the OLNE solution of $\tilde \G := \big( A^i, B^i, \tilde Q^i, R^i: i \in [N] \big)$ satisfies \eqref{Eqn: FBNE, Riccati, 1, initialization}-\eqref{Eqn: FBNE, Riccati, 4, computing Zt}, with each $Q^i, Z_t^i, K_t^i, F_t$ replaced by $\tilde Q^i, \tilde Z_t^i, \tilde K_t^i, \tilde F_t$, respectively. In other words, $\tilde K_t^i = \tilde L_t^i$ and $\tilde F_t = \tilde \Lambda_t^{-1} A$, $\forall \ i \in [N], t \in [T]$.
\end{lemma}

\begin{proof}
By definition, $\tilde M_{T+1}^i = \tilde Q^i$, which confirms \eqref{Eqn: FBNE, Riccati, 1, initialization}. Next, from \eqref{Eqn: OLNE, Riccati, 2, computing Lambda t} and \eqref{Eqn: OLNE, Riccati, 3, computing Lt}, we have $\forall \ i \in [N], t \in [T]$:\vspace{-0.4em}
\begin{equation*}
\begin{aligned}
    &R^i \tilde L_t^i + \hat B^{i\top} \tilde M_{t+1}^i \sum_{j \in [N]} \hat B^j \tilde L_t^j \\
    =  &\hat B^{i\top} \tilde M_{t+1}^i \tilde \Lambda_t^{-1} A \\
    &\hspace{5mm}
    + \hat B^{i\top} \tilde M_{t+1}^i \hspace{-1mm}\sum_{j \in [N]}\hspace{-1mm} \hat B^j (R^j)^{-1} \hat B^{j\top} \tilde M_{t+1}^j \tilde \Lambda_t^{-1} A \\
    =  &\hat B^{i\top} \tilde M_{t+1}^i \Big( I + \sum_{j \in [N]} \hat B^j (R^j)^{-1} \hat B^{j\top} \tilde M_{t+1}^j \Big) \tilde \Lambda_t^{-1} A \\
    =  &\hat B^{i\top} \tilde M_{t+1}^i A,
\end{aligned}\vspace{-0.4em}
\end{equation*}
so $\tilde L_t$ and $\tilde M_{t+1}^i$ satisfy \eqref{Eqn: FBNE, Riccati, 2, computing Kt}. Set $\tilde F_t := A - \sum_{j \in [N]} \hat B^j \tilde L_t^j$ so as to satisfy \eqref{Eqn: FBNE, Riccati, 3, computing Ft}. We proceed to verify \eqref{Eqn: FBNE, Riccati, 4, computing Zt}. First, $\forall \ t \in [T]$:\vspace{-0.4em}
\begin{align} \label{Eqn: 2nd lemma in Jingqi notes, 1}
    \tilde \Lambda_t \tilde F_t &= \Big( I + \sum_{i\in[N]} \hat B^i (R^i)^{-1} \hat B^{i\top} \tilde Z_{t+1}^i \Big) \tilde F_t \\ \label{Eqn: 2nd lemma in Jingqi notes, 2}
    &= \tilde F_t + \sum_{i\in[N]} \hat B^i (R^i)^{-1} R^i \tilde K_t^i \\ \nonumber
    &= \tilde F_t + \sum_{i\in[N]} \hat B^i \tilde K_t^i \\ \label{Eqn: 2nd lemma in Jingqi notes, 3}
    &= A,
\end{align}
where \eqref{Eqn: 2nd lemma in Jingqi notes, 2} follows from \eqref{Eqn: 1st lemma in Jingqi notes}, while \eqref{Eqn: 2nd lemma in Jingqi notes, 3} follows from \eqref{Eqn: FBNE, Riccati, 3, computing Ft}. Next, by rearranging \eqref{Eqn: FBNE, Riccati, 2, computing Kt}, we obtain that:\vspace{-0.4em}
\begin{equation} 
\label{Eqn: 1st lemma in Jingqi notes}
    R^i \tilde K_t^i = \hat B^{i\top} \tilde Z_{t+1}^i \Big( A_t - \sum_{j \in [N]} \hat B^j \tilde K_t^j \Big) = \hat B^{i \top} \tilde Z_{t+1}^i \tilde F_t.\vspace{-0.5em}
\end{equation}
Moreover:\vspace{-0.4em}
\begin{align} \nonumber
    &\tilde Q^i + \tilde F_t^\top \tilde M_{t+1}^i \tilde F_t + \tilde L_t^{i\top} R^i \tilde L_t^i \\ \label{Eqn: 3rd lemma in Jingqi notes, 1}
    = \hspace{0.5mm} &\tilde Q^i + \tilde F_t^\top \tilde M_{t+1}^i \tilde F_t + \tilde L_t^{i\top} \hat B^{i\top} \tilde M_{t+1}^i \tilde F_t \\ \label{Eqn: 3rd lemma in Jingqi notes, 2}
    = \hspace{0.5mm} &\tilde Q^i + \Big(\tilde F_t + \sum_{j \in [N]} \hat B^j \tilde L_t^j \Big)^\top \tilde M_{t+1}^i \tilde F_t \\ \label{Eqn: 3rd lemma in Jingqi notes, 3}
    = \hspace{0.5mm} &\tilde Q^i + A^\top \tilde M_{t+1}^i \tilde F_t \\ \label{Eqn: 3rd lemma in Jingqi notes, 4}
    = \hspace{0.5mm} &\tilde Q^i + A^\top \tilde M_{t+1}^i \tilde \Lambda_t^{-1} A \\ \label{Eqn: 3rd lemma in Jingqi notes, 5}
    = \hspace{0.5mm} &\tilde M_t^i,
\end{align}
so $\tilde L_t$ and $\tilde M_{t+1}^i$ satisfy \eqref{Eqn: FBNE, Riccati, 3, computing Ft}. Above, \eqref{Eqn: 3rd lemma in Jingqi notes, 1} follows from \eqref{Eqn: 1st lemma in Jingqi notes}, \eqref{Eqn: 3rd lemma in Jingqi notes, 2} follows from Remark \ref{Remark: Mti only has nonzero ith row blocks}, \eqref{Eqn: 3rd lemma in Jingqi notes, 3} follows from \eqref{Eqn: FBNE, Riccati, 3, computing Ft}, \eqref{Eqn: 3rd lemma in Jingqi notes, 4} follows from \eqref{Eqn: 2nd lemma in Jingqi notes, 1}-\eqref{Eqn: 2nd lemma in Jingqi notes, 3}, and \eqref{Eqn: 3rd lemma in Jingqi notes, 5} follows from \eqref{Eqn: OLNE, Riccati, 4, computing Mt}.
\end{proof}

Recall that Lemmas \ref{Lemma: 4th lemma in Jingqi notes} and \ref{Lemma: 3rd lemma in Jingqi notes} respectively imply that, loosely speaking, the OLNE of $\G$ and of $\tilde \G$ coincide, and the OLNE of $\tilde \G$ satisfies the Riccati equations \eqref{Eqn: FBNE, Riccati, 1, initialization}-\eqref{Eqn: FBNE, Riccati, 4, computing Zt} for $\tilde \G$. Together, Lemmas~\ref{Lemma: 4th lemma in Jingqi notes} and \ref{Lemma: 3rd lemma in Jingqi notes} establish that the OLNE trajectory of $\G$ matches the FBNE trajectory of $\tilde \G$ (Lemma \ref{Lemma: OLNE of G equals FBNE of tilde G}). Thus, if the FBNE of $\tilde \G$ and of $ \G$ coincide, then the OLNE and FBNE trajectories of $\G$ are aligned (Thm. \ref{Thm: OLNE vs FBNE of G}).

\begin{lemma} \label{Lemma: OLNE of G equals FBNE of tilde G}
If Assumption~\ref{Assumption: LQ Games} holds and $\Lambda_t^{-1}$ exists $\forall t$, then the OLNE of $\G$ satisfies \eqref{Eqn: FBNE, Riccati, 1, initialization}-\eqref{Eqn: FBNE, Riccati, 4, computing Zt} for $\tilde \G$,~and~$\Lambda_t^{-1}A = \tilde F_t$,~$\forall t$. 
\end{lemma}

\begin{theorem} \label{Thm: OLNE vs FBNE of G}
Consider an LQ game $\G$ and its auxiliary game $\tilde \G$. Under Assumption~\ref{Assumption: LQ Games}, suppose $\Lambda_t^{-1}$ exists $\forall t$ and $\tilde K_t^i = K_t^i$ $\forall i \in [N], t \in [T]$, then $F_t = \Lambda_t^{-1} A$, $\forall t$ and the FBNE and OLNE trajectories of $\G$ are identical from any initial state.
\end{theorem}



\section{Bounding the FBNE-to-OLNE Gap of LQ Games}
\label{sec: Bounding the Gap Between the FBNE and OLNE of LQ Games}

Given an LQ game $\G := (A^i, B^i, Q^i, R^i: i \in [N])$, Thm. \ref{Thm: OLNE vs FBNE of G} suggests that the FBNE-to-OLNE gap of $\G$ can be characterized using the difference between the FBNE of $\G$, i.e., $(K_t^i: i \in [N], t \in [T])$, and of its auxiliary game $\tilde \G := (A^i, B^i, \tilde Q^i, R^i: i \in [N])$, i.e., $(\tilde K_t^i: i \in [N], t \in [T])$, as quantified by $\Vert \tilde K_t - K_t \Vert_2$. Below, Prop. \ref{Prop: Diff in K in terms of Diff in Q} complements Thm. \ref{Thm: OLNE vs FBNE of G} by upper bounding the deviation of the FBNE strategy of a dynamic LQ game when its state cost matrices are perturbed. 
Together, Thm. \ref{Thm: OLNE vs FBNE of G} and Prop. \ref{Prop: Diff in K in terms of Diff in Q} upper bound the the FBNE-to-OLNE gap of $\G$, as summarized in Thm.~\ref{Thm: FBNE and OLNE of G, using diff between Q and tilde Q}. 


\begin{proposition} \label{Prop: Diff in K in terms of Diff in Q}
Given $\G := (A^i, B^i, Q^i, R^i: i \in [N])$ and $\hat \G := (A^i, B^i, \hat Q^i, R^i: i \in [N])$, let $\epsilon := \max_{i \in [N]} \Vert \hat Q^i - Q^i \Vert_2$, and let $(\hat{Z}_t^i, \hat{K}_t^i, \hat{F}_t, \hat{P}_t, \hat{S}_t: i\in[N], t\in[T])$ be the solutions of \eqref{Eqn: FBNE, Riccati, 1, initialization}-\eqref{Eqn: Compact version of FBNE, Riccati, 2, computing Kt} for $\hat \G$. Set $\delta \hat P_t := \Vert \hat P_t - P_t \Vert_2$, $\delta \hat S_t := \Vert \hat S_t - S_t \Vert_2$, $\delta \hat K_t := \Vert \hat K_t - K_t \Vert_2$, $\delta \hat Z_t^i := \Vert \hat Z_t^i - Z_t^i \Vert_2$ $\forall \ t \in [T], i \in [N]$. If $\delta \hat P_t < 1/\Vert P_t^{-1} \Vert_2$ $\forall \ t \in [T]$, the following recursive upper bounds for $\delta \hat P_t$, $\delta \hat S_t$, $\delta \hat K_t$, $\delta \hat Z_t^i$ hold:\vspace{-0.4em}
\begin{align} \label{Eqn: delta Z T+1 i}
    \delta \hat Z_{T+1}^i &= \epsilon, \ \forall i \in [N], \\
    \label{Eqn: delta Pt}
    \delta \hat P_t &\leq \sqrt{N} \ \Vert B \Vert_2^2 \ \max_{i \in [N]} \delta \hat Z_{t+1}^i, \\ \label{Eqn: delta St}
    \delta \hat S_t &\leq \sqrt{N} \ \Vert A \Vert_2 \Vert B \Vert_2 \max_{i \in [N]} \delta \hat Z_{t+1}^i \\ \label{Eqn: delta Kt}
    \delta \hat K_t &\leq \frac{\Vert P_t^{-1} \Vert_2}{1 - \Vert P_t^{-1} \Vert_2 \delta \hat P_t} \cdot \big( \Vert K_t \Vert_2 \delta \hat P_t + \delta \hat S_t \big) \\ \label{Eqn: delta Zti}
    \delta \hat Z_t^i &\leq \big(\Vert K_t^i \Vert_2^2 + 1 \big) \epsilon + \Vert B \Vert_2 \big( \Vert Z_{t+1}^i \Vert_2 + \delta \hat Z_{t+1}^i \big) \\ \nonumber
    &\hspace{1cm} \cdot \big( 2 \Vert A - B K_t \Vert_2 + \Vert B \Vert_2 \delta \hat K_t \big) \ \delta \hat K_t \\ \nonumber
    &\hspace{5mm} + \Vert A - B K_t \Vert_2^2 \cdot \delta \hat Z_{t+1}^i \\ \nonumber
    &\hspace{5mm} + (\Vert R^i \Vert_2 + \epsilon) (2 \Vert K_t^i \Vert_2 + \delta \hat K_t) \delta \hat K_t.
\end{align}
\end{proposition}

\begin{proof}
The proof follows by carefully applying known results from the perturbation theory of linear equations (\cite{Demmel1997AppliedNumericalLinearAlgebra}, Section 2.2), as well as the triangle and Cauchy-Schwarz inequalities, to the FBNE Riccati equations \eqref{Eqn: FBNE, Riccati, 1, initialization}-\eqref{Eqn: FBNE, Riccati, 4, computing Zt}, as well as \eqref{Eqn: Compact version of FBNE, Riccati, 2, computing Kt}. Details are omitted for brevity.
\end{proof}

The following example shows 
the theoretical upper bound for $\delta \hat K_t$ in \eqref{Eqn: delta Z T+1 i}-\eqref{Eqn: delta Zti} is nearly tight for some 
$\G$ and $\hat \G$.

\begin{figure}
    \hspace{-0.5cm}
    \centering
    \includegraphics[width=0.7\linewidth]{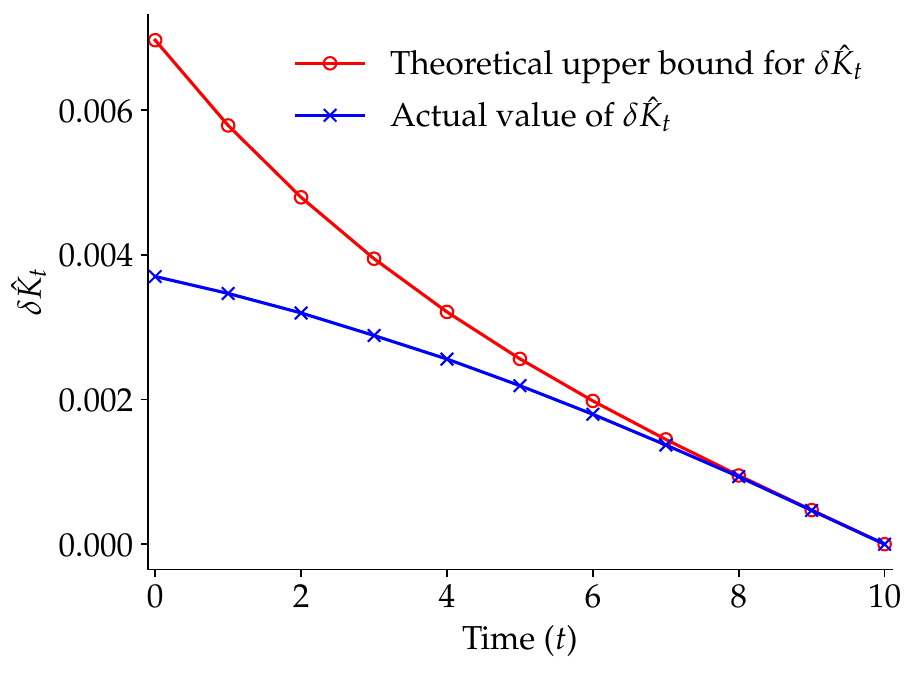}\vspace{-1em}
    \caption{The actual values of $\delta \hat K_t$ 
    and the theoretical upper bound of $\delta \hat K_t$ as given by \eqref{Eqn: delta Z T+1 i}-\eqref{Eqn: delta Zti} (red, with circle marks), corresponding to the LQ games $\G$ and $\hat \G$ described in Example \ref{Ex: Tightness of Upper Bound}. The theoretical upper bound for $\delta \hat K_t$ closely matches the actual value of $\delta \hat K_t$ throughout the first few iterates of the backward iteration process, i.e., when $t$ is close to $T = 10$. Significant divergence between the theoretical upper bound and actual value of $\delta \hat K_t$ only occurs near the start of the time horizon.}\vspace{-1.6em}
    \label{fig: K_error_gap_theory_empirical}
\end{figure}

\begin{example} \label{Ex: Tightness of Upper Bound}
Consider 2-agent, 10-stage LQ games $\G := (A^i, B^i, Q^i, R^i: i \in \{1, 2\})$ and $\hat \G := (A^i, B^i, \hat Q^i, R^i: i \in \{1, 2\})$, with $A^i = B^i = 1$, $R^i = 300$, $\forall \ i \in \{1, 2\}$, $Q^i = I_{2 \times 2}$ (the $2 \times 2$ identity matrix), for each $i \in [2]$, and: \vspace{-0.4em}
\begin{equation*}
    \hat Q^1 = 1.1 \ I_{2 \times 2}, \hspace{5mm} \hat Q^2 = \begin{bmatrix}
        1 & 0.1 \\
        0.1 & 1
    \end{bmatrix}.\vspace{-0.4em}
\end{equation*}
Fig. \ref{fig: K_error_gap_theory_empirical} plots the theoretical upper bound (from \eqref{Eqn: delta Z T+1 i}-\eqref{Eqn: delta Zti}) and actual values of $\delta \hat K_t$, i.e., the FBNE-to-OLNE gap of $\G$ and $\hat \G$. The theoretical upper bound for $\delta \hat K_t$ closely tracks the $\delta \hat K_t$ value during initial steps of the backward recursion from $T = 10$, i.e., when $t$ is close to $T$.
\end{example}

The following result follows directly from Prop. \ref{Prop: Diff in K in terms of Diff in Q}.

\begin{theorem} \label{Thm: FBNE and OLNE of G, using diff between Q and tilde Q}
Given an LQ game $\G := (A^i, B^i, Q^i, R^i: i \in [N])$ and its auxiliary game $\tilde \G:= (A^i, B^i, \tilde Q^i, R^i: i \in [N])$, 
let $\epsilon := \max_{i \in [N]} \Vert \tilde Q^i - Q^i \Vert_2$. Then the FBNE-to-OLNE gap of $\G$ can be bounded as:
\begin{equation*}
    \Vert L_t - K_t \Vert_2 \leq \frac{\Vert P_t^{-1} \Vert_2}{1 - \Vert P_t^{-1} \Vert_2 \delta \tilde P_t} \cdot \big( \Vert K_t \Vert_2 \delta \tilde P_t + \delta \tilde S_t \big),\vspace{-0.2em}
\end{equation*}
in conjunction with variants of \eqref{Eqn: delta Z T+1 i}-\eqref{Eqn: delta St}, \eqref{Eqn: delta Zti} with $\delta \tilde K_t := \Vert \tilde K_t -  K_t \Vert_2 = \Vert L_t -  K_t \Vert_2$, $\delta \tilde S_t := \Vert \tilde S_t - S_t \Vert_2$, $\delta \tilde P_t := \Vert \tilde P_t - P_t \Vert_2$, $\delta \tilde Z_t^i := \Vert \tilde Z_t^i - Z_t^i \Vert_2$, $\forall \ t \in [T], i \in [N]$, where $L_t \in \R^{m \times n}$ is defined block-wise by $[L_t]_{i,:} = L_t^i, \forall \ i \in [N]$.
\end{theorem}

\section{Simulation Studies}
\label{sec: Simulation Studies}



We present a Monte Carlo study to validate the sufficient condition in Thm.~\ref{Thm: OLNE vs FBNE of G} for the alignment of the FBNE and OLNE of dynamic LQ games. 
We then study the FBNE-to-OLNE gaps of two specific games from our Monte Carlo study that violate the sufficient condition to varying degrees of severity. Our observations confirm our conclusions in Thm. \ref{Thm: FBNE and OLNE of G, using diff between Q and tilde Q} that the more severely a dynamic LQ game violates the sufficient condition $\delta \tilde K_t = 0$, the more significantly its FBNE and OLNE solutions diverge.

\begin{figure}[!t]
    \includegraphics[width=1.0\linewidth]{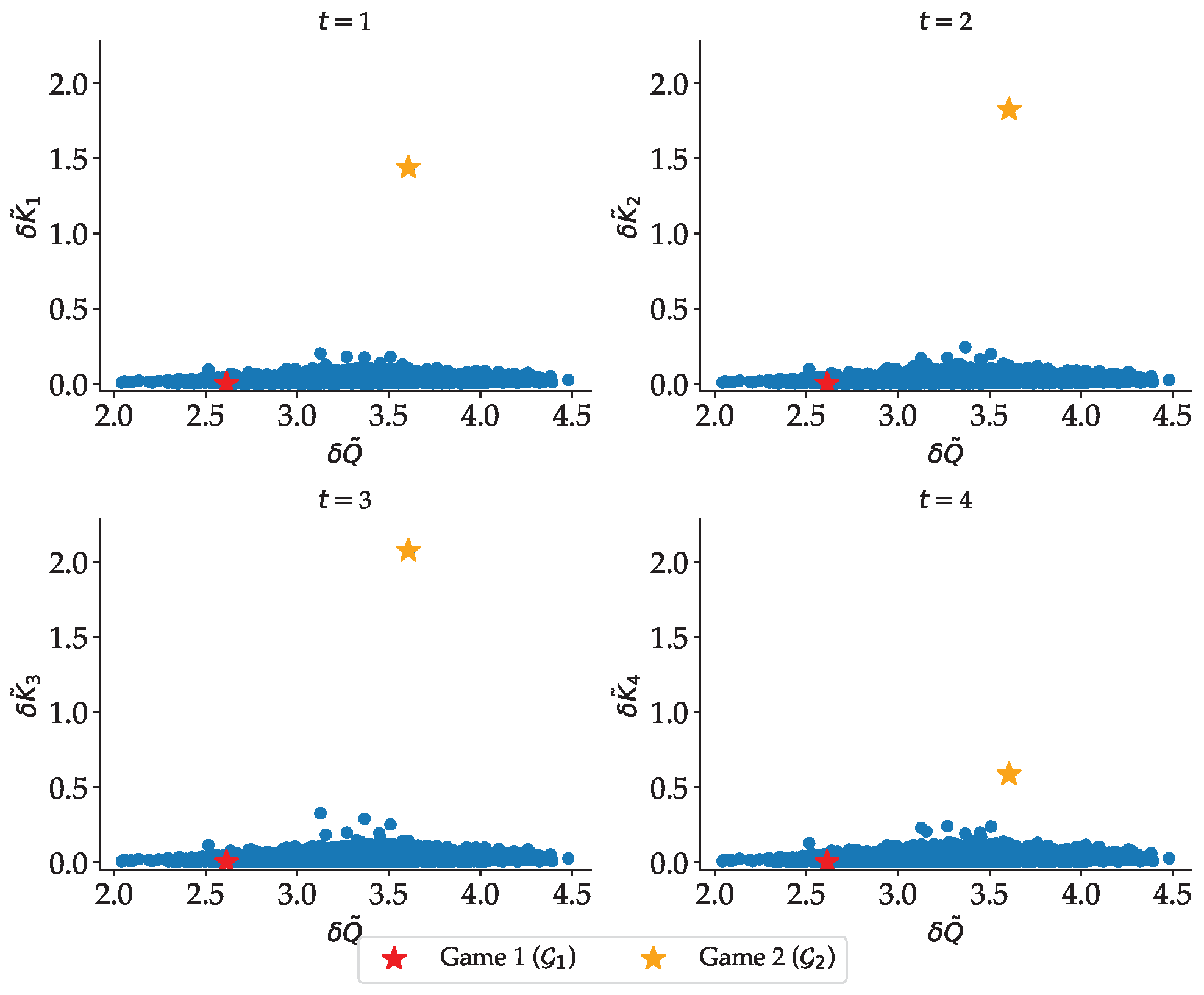}
    \vspace{-2.0em}
    \caption{
    Scatter plot of $\delta \tilde K := \Vert \tilde K_t - K_t \Vert_2$ vs. $\delta \tilde Q := \Vert \tilde Q - Q \Vert_2$, at each $t \in [4]$, for each of the 10000 sampled LQ dynamic games. 
    Each sampled game admits a \textbf{unique} FBNE and a \textbf{unique} OLNE. These samples violate to different degrees the sufficient condition $\delta \tilde K_t = 0$ for the FBNE and OLNE trajectories to be aligned (Thm. \ref{Thm: OLNE vs FBNE of G}). Specifically, the game $\G_1$ barely violates the condition, while the game $\G_2$ violates the condition much more severely. Moreover, many sampled games correspond to large values of $\delta \tilde Q$ but tiny values of $\delta \tilde K_t$, which shows the difficulty of obtaining nontrivial lower bounds for $\delta \tilde K_t$ in terms of $\delta \tilde Q$.
    }\vspace{-1.3em}
    \label{fig:monte-carlo}
\end{figure}

In our Monte Carlo study, we generated 10000 instances of a 2-agent dynamic LQ game with $T = 10$ by independently sampling state cost matrices $Q^i$ while holding all other game parameters fixed at the following values:\vspace{-0.5em}
\begin{equation*}
    A^1 = A^2 = \begin{bmatrix}
        0 & 1 \\ -1 & -1
    \end{bmatrix}, B^1 = B^2 = \begin{bmatrix}
        0 \\ 1
    \end{bmatrix}, R^1 = 3, R^2 = 2.\vspace{-0.3em}
\end{equation*}
Fig. \ref{fig:monte-carlo} presents $\delta \tilde K_t$ vs. $\delta \tilde Q$ scatter plots for all sampled games, for $t \in [4]$. Many of these sampled games correspond to large values of $\delta \tilde Q$ and very small values of $\delta \tilde K_t$, showcasing the difficulty of obtaining a uniform lower bound.

To study the sufficient condition in Thm. \ref{Thm: OLNE vs FBNE of G} in more detail, we sample two games that violate the sufficient condition of the game to different degrees of severity. Without loss of generality, we reindex the game instance marked by the red star in Fig.~\ref{fig:monte-carlo} as $\G_1$, which has state cost matrices given by:\vspace{-0.4em}
\begin{equation*}
\begin{aligned}
Q^1 &= \begin{bmatrix}
1.86 & 0.85 & 0.57 & 0.44 \\
0.85 & 1.99 & 0.35 & 0.25 \\
0.57 & 0.35 & 1.53 & 0.70 \\
0.44 & 0.25 & 0.70 & 1.07
\end{bmatrix}, \\
Q^2 &= \begin{bmatrix}
1.29 & 0.92 & 0.98 & 0.57 \\
0.92 & 1.16 & 0.78 & 0.53 \\
0.98 & 0.78 & 1.63 & 0.53 \\
0.57 & 0.53 & 0.53 & 1.84
\end{bmatrix}.
\end{aligned}\vspace{-0.4em}
\end{equation*}
We observe from Fig. \ref{fig:monte-carlo} that $\G_1$ only barely violates the sufficient condition. Next, we denote by $\G_2$ the game marked by the orange star in Fig. \ref{fig:monte-carlo}, with the state cost matrices:\vspace{-0.4em}
\begin{equation*}
\begin{aligned}
    Q^1 &= \begin{bmatrix}
    1.48 & 1.41 & 1.09 & 0.94 \\
    1.41 & 1.66 & 0.89 & 1.46 \\
    1.09 & 0.89 & 1.64 & 0.87 \\
    0.94 & 1.46 & 0.87 & 1.95
    \end{bmatrix}, \\
    Q^2 &= \begin{bmatrix}
    1.83 & 0.44 & 1.43 & 0.81 \\
    0.44 & 1.97 & 0.63 & 0.87 \\
    1.43 & 0.63 & 1.29 & 0.85 \\
    0.81 & 0.87 & 0.85 & 0.73
    \end{bmatrix}.
\end{aligned}
\end{equation*}
We observe from Fig. \ref{fig:monte-carlo} that $\G_2$ violates the sufficient condition much more severely compared to $\G_1$.



\begin{figure}
    \centering
\includegraphics[width=\linewidth]{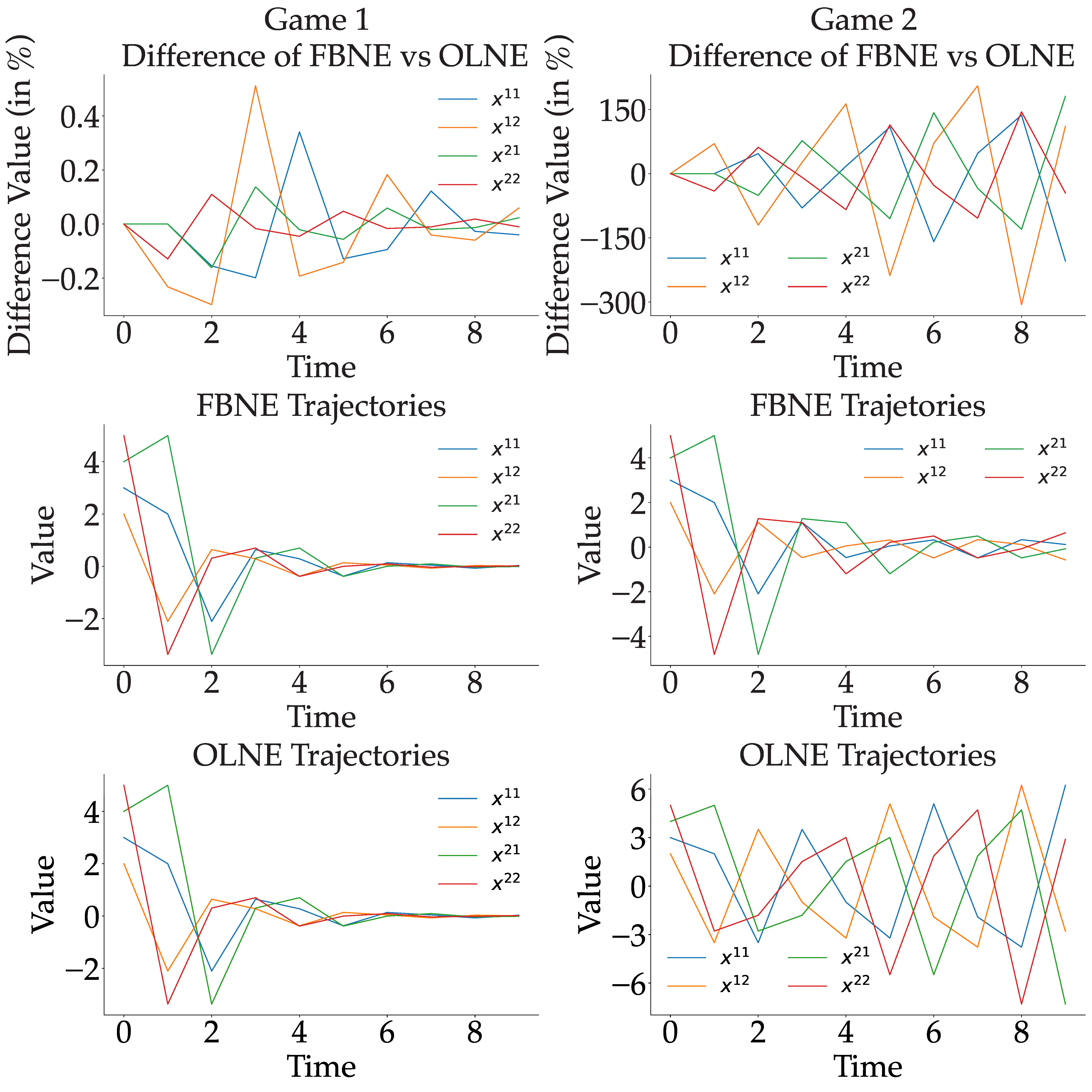}\vspace{-1em}
    \caption{Plots, for $\G_1$ and $\G_2$, of the FBNE trajectories, OLNE trajectories, and the differences between the FBNE and OLNE trajectories. We observe that the FBNE and OLNE trajectories of $\G_1$, which only slightly violates the sufficient condition in Thm. \ref{Thm: OLNE vs FBNE of G} (i.e., 
    $\delta \tilde K_t = 0$), are within 0.25\% of each other. In contrast, $\G_2$ violates the sufficient condition more severely, and its FBNE and OLNE trajectories differ up to 300\%.}\vspace{-1.5em}
    \label{fig:two-agent-traj}
\end{figure}

Fig. \ref{fig:two-agent-traj} plots the FBNE and OLNE trajectories of $\G_1$ and of $\G_2$, as well as the percent difference between the FBNE and OLNE of $\G_1$ and of $\G_2$, normalized with respect to the respective initial state values. We observe that the FBNE and OLNE trajectories for $\G_1$ are within 0.5\% of each other since $\G_1$ only slightly violates the sufficient condition. In contrast, $\G_2$, which violates the sufficient condition by a wider margin, has FBNE and OLNE trajectories that differ up to 300\%. 
Our results illustrate that the degree to which a dynamic LQ game violates the sufficient condition heavily influences how much its FBNE and OLNE trajectories diverge\footnote{\revision{For plots of the FBNE-to-OLNE gap over a wider selection of games, please see the Appendix in the ArXiv version of this work \cite{ChiuLiBhattMehr2024FBNEvsOLNE}.}}.

\section{Conclusion}\vspace{-0.3em}
\label{sec: Conclusion and Future Work}

We presented a principled approach for quantifying the difference between the FBNE and OLNE of a dynamic LQ game. Given a dynamic LQ game $\G$, we constructed an \textit{auxiliary} game $\tilde \G$, a copy of $\G$ with modified state costs, and proved that the OLNE of $\G$ can be synthesized by solving for the FBNE of $\tilde \G$.
We then quantified the FBNE-to-OLNE gap of $\G$ by bounding the difference between the FBNE strategies of $\G$ and $\tilde \G$, i.e., $\Vert K_t - \tilde K_t \Vert_2$, using the difference between the state cost matrices of $\G$ and $\tilde \G$. We presented numerical results that validate our theoretical claims and illustrate the difficulty of obtaining matching lower bounds. Our results provide insight into the FBNE-to-OLNE gap for dynamic games, and facilitate the selection of appropriate equilibrium solutions, with promising applications in areas such as autonomous driving and human-robot interaction.




\printbibliography

\newpage
\appendices
\section{Additional Simulation Results}

\subsection{Heterogenity among Agent Dynamics}
\label{subsec: Heterogenity in Agent Dynamics}

In order to study whether the magnitude of the FBNE-to-OLNE gap of a dynamic LQ game is affected by heterogeneity in agents' dynamics, as encoded by the matrices $\{A^i, B^i: i \in [N]\}$, we randomly sample 10,000 distinct values of $A^1,A^2,B^1,B^2,Q^1$ and $Q^2$ while fixing $R^1 = 3$ and $R^2 = 2$. 

\begin{figure}[h]
    \includegraphics[width=1.0\linewidth]{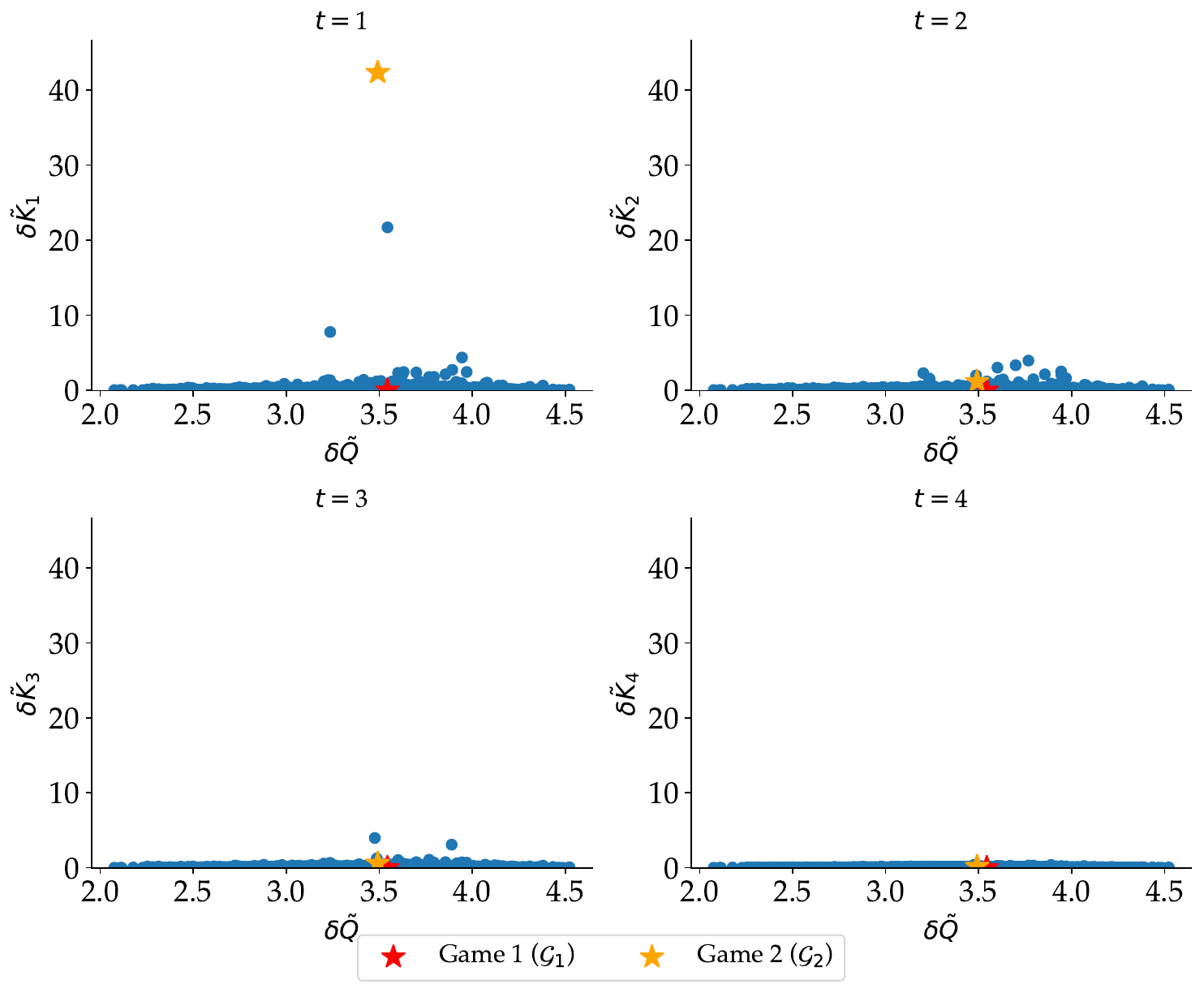}
    \vspace{-2.4em}
    \caption{
    Scatter plot of $\delta \tilde K := \Vert \tilde K_t - K_t \Vert_2$ vs. $\delta \tilde Q := \Vert \tilde Q - Q \Vert_2$, at each $t \in [4]$, for each of the 10000 sampled LQ dynamic games. The dynamics matrices $A^1,A^2,B^1,B^2$ and state cost matrices $Q^1, Q^2$ of each sampled game is independently sampled, while the control cost matrices $R^1$ and $R^2$ are fixed across sampled games.
    }
    \label{fig:monte-carlo-random-A-B}
\end{figure}

Fig. \ref{fig:monte-carlo-random-A-B} presents $\delta \tilde K_t$ vs. $\delta \tilde Q$ scatter plots for all sampled games, at each $t \in [T]$ with $T = 4$.
The corresponding values of $A^1,A^2,B^1,B^2,Q^1$ and $Q^2$ for $\G_3$ and $\G_4$ are as follows.
\begin{itemize}
    \item \textbf{Game $\G_3$}:
    \begin{equation*}
    A^1 = \begin{bmatrix}
        0.39 & 0.87 \\ 0.15 & 0.37
    \end{bmatrix}, 
    A^2 = \begin{bmatrix}
        0.07 & 0.38 \\ 0.04 & 0.30
    \end{bmatrix},
    \end{equation*}
    \begin{equation*}
        B^1 = \begin{bmatrix}
        0.78 \\ 0.97
    \end{bmatrix}, B^2 = \begin{bmatrix}
        0.14 \\ 0.001
    \end{bmatrix},
\end{equation*}
\[
Q_1 = \begin{bmatrix}
1.88 & 0.67 & 0.49 & 0.75 \\
0.67 & 1.88 & 1.03 & 1.81 \\
0.49 & 1.03 & 1.58 & 0.70 \\
0.75 & 1.81 & 0.70 & 1.98 \\
\end{bmatrix},
\]

\[
Q_2 = \begin{bmatrix}
0.72 & 0.54 & 0.58 & 0.68 \\
0.54 & 1.20 & 0.79 & 0.39 \\
0.58 & 0.79 & 1.53 & 0.26 \\
0.68 & 0.39 & 0.26 & 1.71 \\
\end{bmatrix}
\]

\item \textbf{Game $\G_4$}
\begin{equation*}
    A_1 = \begin{bmatrix}
0.16 & 0.96 \\
0.96 & 0.55 \\
\end{bmatrix}, A_2 = \begin{bmatrix}
0.65 & 0.94 \\
0.29 & 0.81 \\
\end{bmatrix},
\end{equation*}
\begin{equation*}
    B_1 = \begin{bmatrix}
0.89 \\
0.49 \\
\end{bmatrix}, B_2 = \begin{bmatrix}
0.62 \\
0.56 \\
\end{bmatrix},
\end{equation*}
\[
Q_1 = \begin{bmatrix}
1.04 & 0.32 & 0.80 & 0.66 \\
0.32 & 1.03 & 0.56 & 1.22 \\
0.80 & 0.56 & 1.97 & 1.06 \\
0.66 & 1.22 & 1.06 & 1.79 \\
\end{bmatrix},
\]

\[
Q_2 = \begin{bmatrix}
0.94 & 1.03 & 1.07 & 0.90 \\
1.03 & 1.40 & 1.34 & 1.01 \\
1.07 & 1.34 & 1.35 & 0.88 \\
0.90 & 1.01 & 0.88 & 1.61 \\
\end{bmatrix}.
\]
\end{itemize}

In order to further investigate how the heterogeneity across agents' dynamics affects the FBNE-to-OLNE gap of the dynamic game, we perform a Monte Carlo study of 2000 different games with different values of the following quantities:
\begin{align*}
    &\max_{i\in\{1,2\}} \left\Vert A_i - \text{avg}(A) \right\Vert,
    \hspace{5mm} \max_{i\in\{1,2\}} \left\Vert B_i - \text{avg}(B) \right\Vert,
\end{align*}
where $\text{avg}(A) := \frac{1}{2}(A_1 + A_2)$ and $\text{avg}(B) := \frac{1}{2}(B_1 + B_2)$.

\begin{figure}[h]
    \includegraphics[width=1.0\linewidth]{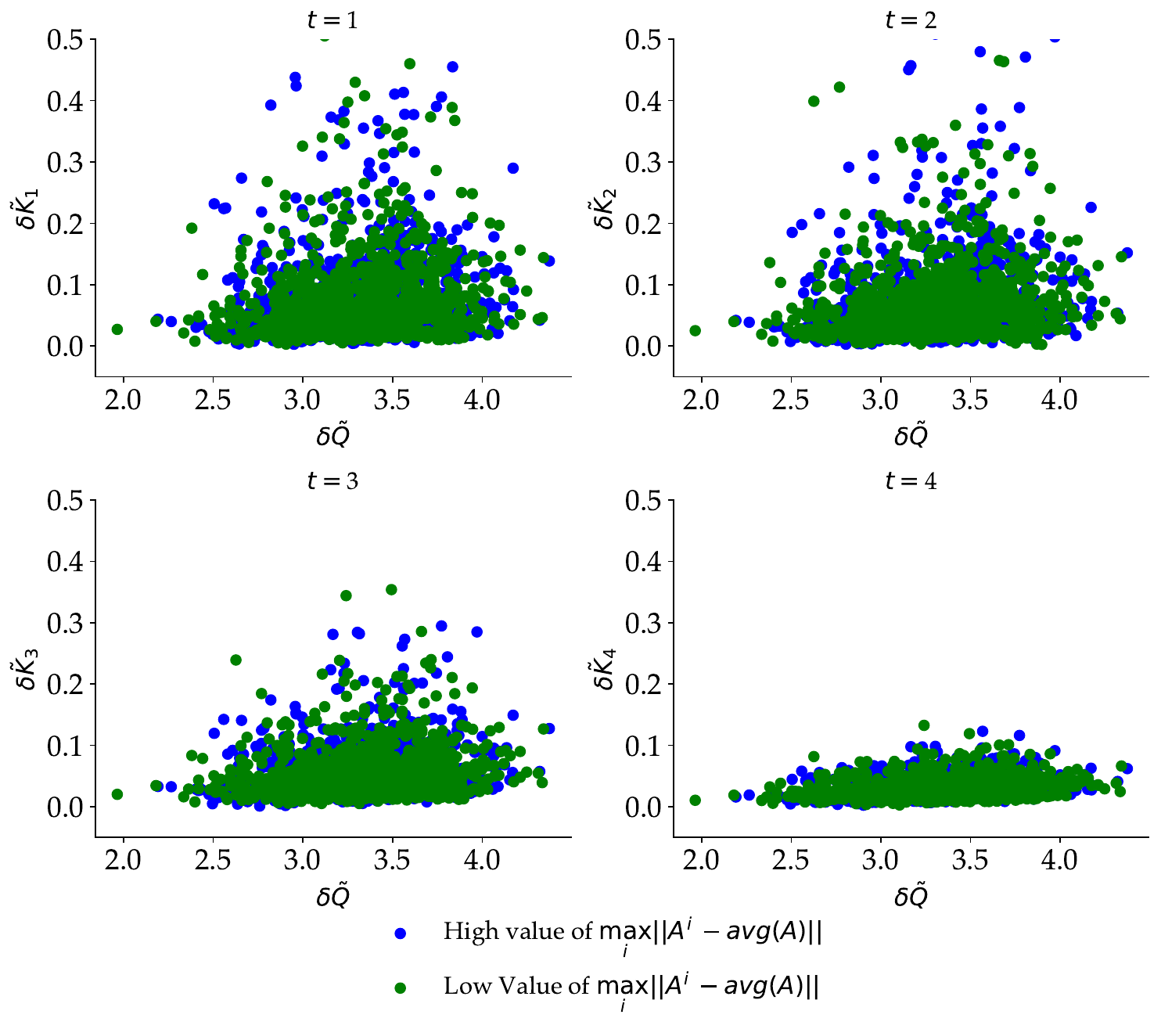}
    \vspace{-2.4em}
    \caption{
    Scatter plot of $\delta \tilde K := \Vert \tilde K_t - K_t \Vert_2$ vs. $\delta \tilde Q := \Vert \tilde Q - Q \Vert_2$, at each $t \in [4]$, for each of the 2000 sampled LQ dynamic games.  Each sample has randomly sampled values of $A^1,A^2,Q^1$ and $Q^2$. 1000 of these samples have high values of $\max_i \Vert A_i - \text{avg}(A) \Vert$ while the rest have low values of $\max_i \Vert A_i - \text{avg}(A) \Vert$, where $\text{avg}(A) := \frac{1}{2}(A_1 + A_2)$.
    }
    \label{fig:monte-carlo-random-A}
\end{figure}

\begin{figure}
    \includegraphics[width=1.0\linewidth]{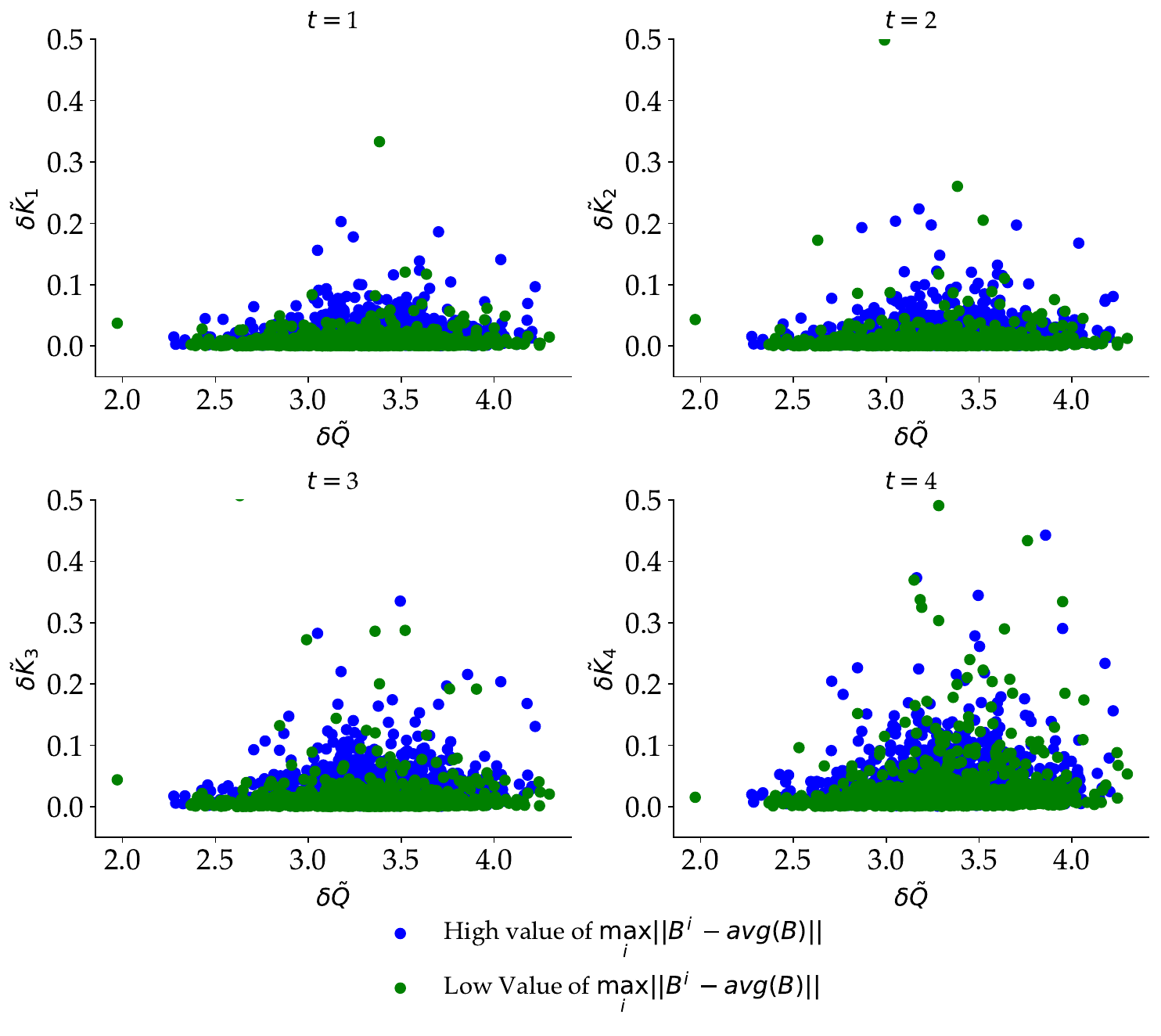}
    \vspace{-2.4em}
    \caption{
    Scatter plot of $\delta \tilde K := \Vert \tilde K_t - K_t \Vert_2$ vs. $\delta \tilde Q := \Vert \tilde Q - Q \Vert_2$, at each $t \in [4]$, for each of the 2000 sampled LQ dynamic games.  Each sample has randomly sampled values of $B^1,B^2,Q^1$ and $Q^2$. 1000 of these samples have high values of $\max_i \Vert B_i - \text{avg}(B) \Vert$ while the rest have low values of $\max_i \Vert B_i - \text{avg}(B) \Vert$, where $\text{avg}(B) := \frac{1}{2}(B_1 + B_2)$.
    }
    \label{fig:monte-carlo-random-B}
\end{figure}

The results of the Monte Carlo study, plotted in Figures~\ref{fig:monte-carlo-random-A} and \ref{fig:monte-carlo-random-B}, reveal that on average, dynamic LQ games with greater heterogeneity between agents' dynamics matrices empirically exhibit larger FBNE-to-OLNE gaps. This result appears to support the hypothesis that heterogeneity among agents' dynamics may widen the gap between the agents' equilibrium behaviors exhibited under open-loop and feedback equilibrium patterns.

\subsection{\texorpdfstring{Dense Sampling Around Game $\G_4$ from Figure~\ref{fig:monte-carlo}}{Dense Sampling Around Game G4 from Figure~\ref{fig:monte-carlo}}}
\label{subsec: Dense Sampling Around Game G4}

\begin{figure}
    \includegraphics[width=1.0\linewidth]{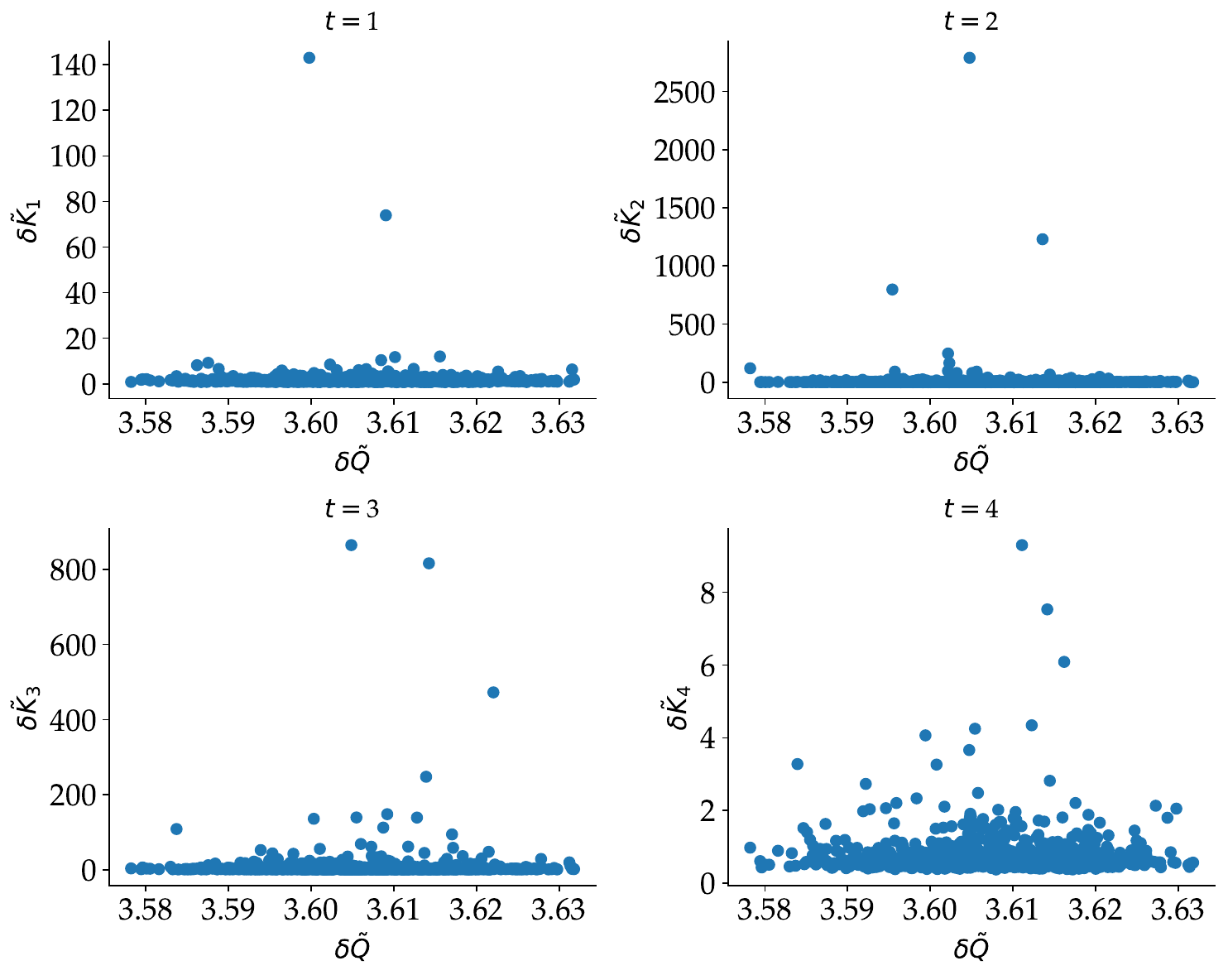}
    \vspace{-2.4em}
    \caption{
    Scatter plot of $\delta \tilde K:= \Vert \tilde K_t - K_t \Vert_2$ vs. $\delta \tilde Q:= \Vert \tilde Q - Q \Vert_2$, at each $t \in [4]$, for each of 1000 densely sampled LQ dynamic games around Game $\G_4$ from Figure~\ref{fig:monte-carlo}. We note that we observe many additional games that, like $\G_4$, have both small values of $\delta \tilde{K}$ and large values of $\delta \tilde{Q}$.
    }
    \label{fig:monte-carlo-dense}
\end{figure}

As Figure~\ref{fig:monte-carlo} illustrates, Game $\G_4$ corresponds to the case when even for small values of $\delta \title{Q}$, we observe high values of $\delta \tilde{K}$ which results in a large FBNE-to-OLNE gap, as quantified by $\delta \tilde{K}$. A natural question is to ask whether Game $\G_4$ is merely an aberration, or whether we can in general expect large FBNE-to-OLNE gaps under small values of $\delta \tilde{Q}$. To answer this question, we perform a Monte Carlo study by densely sampling state cost matrices in the vicinity of the $Q^1$ and $Q^2$ matrices of Game $\G_4$, and plot our findings in Figure~\ref{fig:monte-carlo-dense}. 
We observe that Game $\G_4$ is not simply an exception, and in fact there exist multiple games for which $\delta \tilde{Q}$ is small but $\delta \tilde{K}$ is large. Furthermore, we discover that there exist many games whose $\delta \tilde{K}$ values are significantly higher than those of Game $\G_4$.

\end{document}